\documentclass[aps,prb,twocolumn,showpacs,a4paper,floatfix]{revtex4}
\usepackage{times}
\usepackage[latin1]{inputenc}
\usepackage[english]{babel}
\usepackage[T1]{fontenc}
 \usepackage{graphicx}
\usepackage{epstopdf}
\usepackage[small,FIGTOPCAP]{subfigure}
\usepackage{varioref}
\usepackage{rotating}
\usepackage{mathptmx}
\usepackage{fancyhdr}
\usepackage{amsmath}
\usepackage{amsfonts}
\usepackage{amssymb}
\usepackage{amsbsy}
\usepackage{textcomp}
\usepackage{array}
\usepackage{fixmath}
\usepackage{bm}
\usepackage[active]{srcltx}
\usepackage[colorlinks]{hyperref}
\usepackage[all]{hypcap}

\hypersetup{ pdfcreator=Emiliano Cadelano}

\begin{document}
 \title{Gap opening in graphene by shear strain}
 \author{Giulio Cocco,$^1$ Emiliano Cadelano,$^{1}$ Luciano Colombo$^{1}$}

\email[E-mail me at:]{luciano.colombo@dsf.unica.it}
\affiliation{$^1$Department of Physics, University of Cagliari\\
Cittadella Universitaria,
I-09042 Monserrato (Cagliari), Italy}
\date{\today}

\begin{abstract}

We exploit the concept of strain-induced band structure engineering in graphene through the calculation of its electronic properties under uniaxial, shear, and combined uniaxial-shear deformations. We show that by combining shear deformations to uniaxial strains it is possible modulate the graphene energy gap value from zero up to $0.9$ eV. Interestingly enough, the use of a shear component allows for a gap opening at moderate absolute deformation, safely smaller than the graphene failure strain.
\end{abstract}
\maketitle

Graphene exhibits a number of exotic electronic properties,
such as unconventional integer quantum Hall effect, ultrahigh electron mobility, electron-hole symmetry and ballistic transport even at room temperature.\cite{novomob,neto,novoselov2} Full account of these features is provided by the relativistic Dirac theory\cite{gusynin} suitably developed within the standard condensed matter formalism. A key feature of graphene is that its electronic density of states vanishes at the so-called Dirac points, where  the valence and the conduction bands cross with a linear energy-momentum dispersion. Due to the hexagonal symmetry of graphene, the Dirac points are located at two high-symmetry points of its Brillouin zone.

While many other properties of graphene are very promising for nanoelectronics, its zero-gap semiconductor nature is detrimental, since it prevents the pinch off of charge current as requested in conventional electronic devices.
Different attempts have been therefore tried in order to induce a gap, for instance by quantum confinement of electrons and holes in graphene nanoribbons\cite{han} or quantum dots.\cite{sols}
These patterning techniques are unfortunately affected by the edge roughness problem,\cite{mucciolo} namely: the edges 
are extensively damaged and 
the resulting lattice disorder can even suppress the efficient charge transport.
The sensitivity to the edge structure has been demonstrated through explicit calculations of the electronic states in ribbons.\cite{Nakada}  More recently, it  has been shown experimentally that a band gap as large as $0.45$ eV can be opened if a graphene sheet is placed on an Ir(111) substrate and exposed to patterned hydrogen adsorption.\cite{balog}

Alternatively, an electronic band gap can be obtained by growing graphene sheets on an appropriately chosen substrate, inducing a strain field controllable by temperature.\cite{ni1,shemella,kim,ni2}
Recently, it has been experimentally shown that by using flexible substrates a reversible and controlled strain up to $\sim18$\%\cite{kim} can be generated with measurable variations in the optical, phonon and electronic properties of
graphene.\cite{ni2}
This interesting result suggests that gap opening could be engineered by strain, rather than by patterning. The idea has been theoretically validated by Peirera and Castro Neto\cite{pereira:prb2009} showing that a gap is indeed generated by applying an uniaxial strain as large as $\sim 23$\%, approaching the graphene failure strain $\varepsilon_{f}= 25$\%.\cite{lee,cadelano} This large value stands for the high robustness of the gapless feature of graphene under deformation.
The same Authors propose an alternative origami technique\cite{pereira:prl2009} aimed at generating local strain profiles by means of appropriate geometrical patterns in the substrate, rather than by applying strain directly to the graphene sheet.
\begin{figure}[tp]
\centering
{\includegraphics[width= 0.3\textwidth, angle=0]{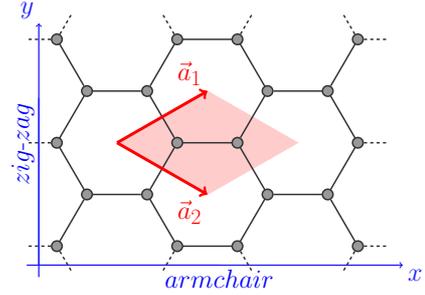}}\\
\caption{(Color online) 
Top view of the hexagonal graphene lattice with its lattice 
vectors $\vec{a}_{1,2}=a_{0}\left(\tfrac{3}{2}, \pm\tfrac{\sqrt{3}}{2}\right) $, where $a_{0}$ is the equilibrium C-C distance. Axis $x$ and $y$ corresponds to the armchair and zig-zag direction, respectively. Shaded area represents the unit cell.
}\label{fig:lattice}
\end{figure}

In this work, we further develop the above concept of strain engineering by showing how a combination of shear and uniaxial strain can be used to open a gap in a range of reversible and more easily accessible deformations, ranging in between $12$\% and $17$\%.
We also discuss the merging of Dirac points,\cite{montambaux} which is involved into the gap opening process.

The electronic structure of graphene has been computed for each deformed configuration by means of a semi-empirical $sp^{3}$ tight-binding (TB) model, making use of the two-center parameterization by Xu \textit{et al.}.\cite{xu} 
Despite its semi-empirical character, the present TB model correctly provides the occurrence of Dirac points in the band structure of graphene in its equilibrium geometry.
Furthermore, the Xu \textit{et al.} parametrization provides accurate scaling functions for the variation of the TB hopping integrals upon lattice distortions. This feature is instrumental for investigating gap opening in graphene by strain. The reliability of the present TB model in describing the strain-related features of graphene has been recently established.\cite{cadelano,cadelano2}

\begin{figure}[tp]
 \centering%
{\includegraphics[width= 0.45\textwidth]{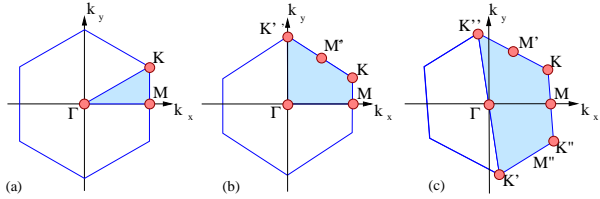}}
\caption{(Color online) Brillouin zone of graphene under strain. The shaded areas are the corresponding irreducible part: (a) undeformed BZ with $6/mmm$ hexagonal symmetries; (b) BZ deformed by uniaxial strain with $mmm$ rhombic symmetry; (c) BZ deformed by shear strain with $2/m$ monoclinic symmetry.
}\label{fig:structure+bz}
\end{figure}
\begin{figure}[tbp]
\centering%
\subfigure[\label{capt:HOMO LUMO  0pc}undeformed graphene]%
{\includegraphics[width= 0.32\textwidth, angle=0]{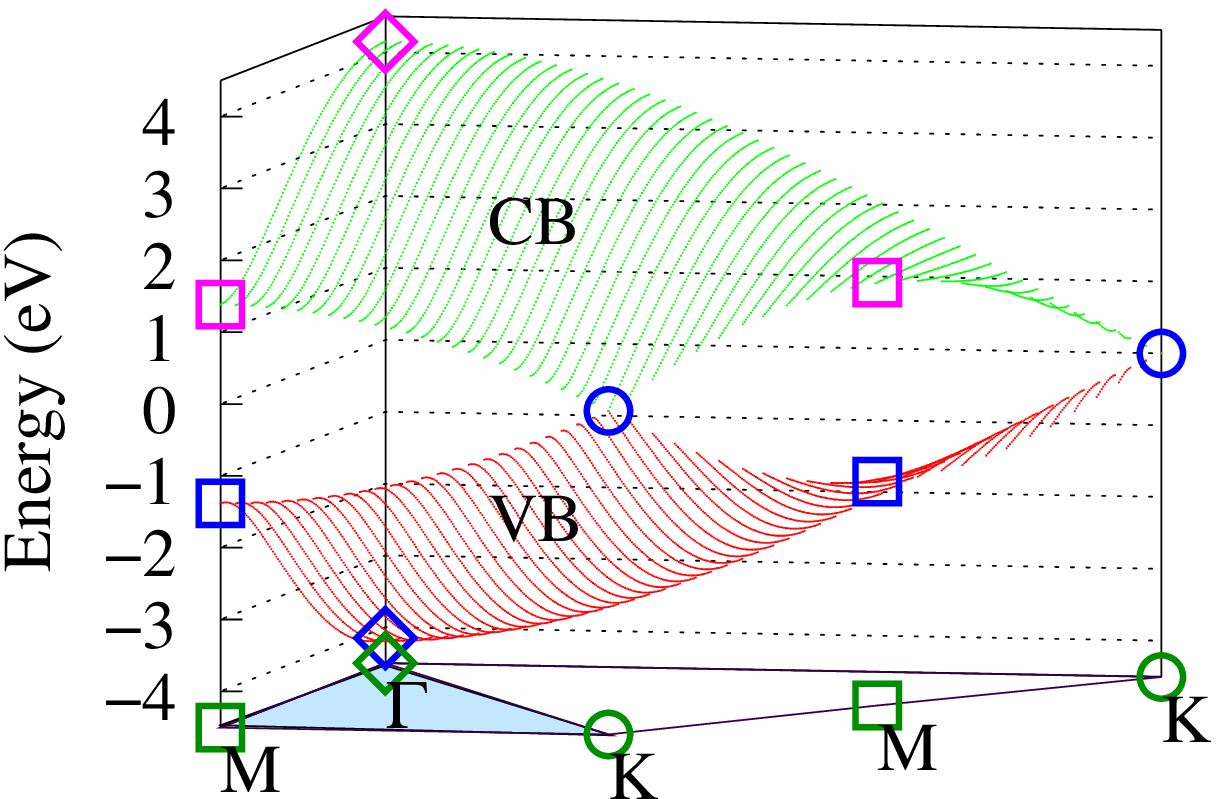}}
\subfigure[\label{capt:HOMO LUMO  arm 15pc} uniaxial (armchair) strain with $\zeta=15$\%]%
{\includegraphics[width= 0.32\textwidth, angle=0]{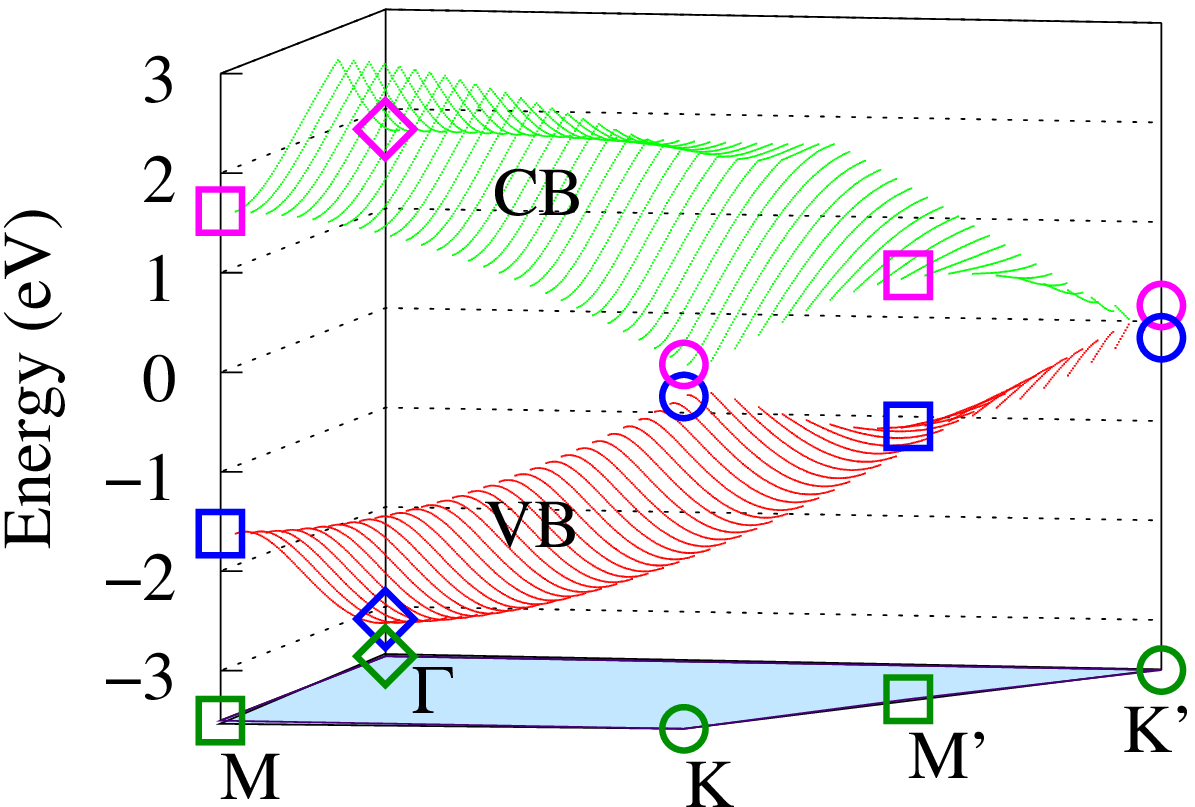}}
\subfigure[\label{capt:HOMO LUMO  zz 15pc} uniaxial (zig-zag) strain with $\zeta=15$\%]%
{\includegraphics[width= 0.32\textwidth, angle=0]{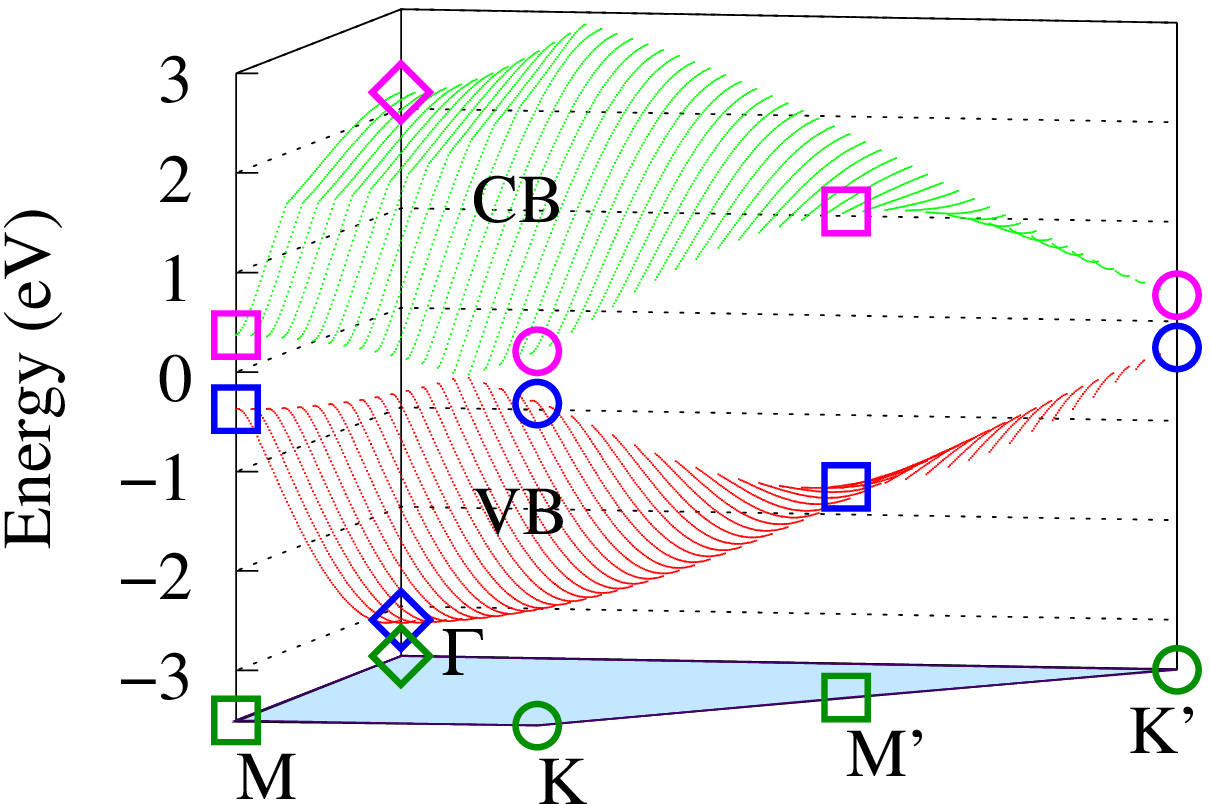}}%
\caption{
(Color online)  Top of the valence band (red, marked as VB) and bottom of the conduction band (green, marked as CB) of graphene under uniaxial strain. Panel a: band structure of the undeformed lattice. Panel b and c: band structure under uniaxial strain along the armchair and  the zig-zag direction, respectively. Symbols connect the high-symmetry points of the BZ (bottom shaded area) to the energy of the corresponding electronic states.}
\label{fig:idro+uniax}
\end{figure}
Graphene is an hexagonal lattice with two carbon atoms per unit cell and a lattice basis defined by
the vectors $\left( \vec{a}_{1},\vec{a}_{2}\right) $, as shown in Fig.\ref{fig:lattice}, with a nearest-neighbor carbon-carbon distance as small as  $a_{0}=~1.42$~\AA. The in-plane elastic behavior of the honeycomb lattice is isotropic in the linear regime, but two inequivalent crystallographic directions can be nevertheless defined: the so-called armchair and zig-zag directions, shown in Fig.\ref{fig:lattice} as $x$ and $y$ axis, respectively.
According to the Cauchy-Born rule, when straining a graphene sample its lattice vectors are affected accordingly, as well as the associated reciprocal vectors $\left( \vec{b}_{1},\vec{b}_{2}\right)$.
The deformed vectors are given by $a'_{i}=\left(\epsilon_{ij}a_{j}+\delta_{ij}a_{j}\right)$, where $\hat{\epsilon}=\left\lbrace \epsilon_{ij}\right\rbrace $ is the strain tensor describing the deformation and $i,j=x,y$. The condition $\vec{a}'_{k}\cdot\vec{b}'_{l}= 2\pi\delta_{kl}$ (where $k,l=1,2$) allows one to obtain the deformed reciprocal lattice vectors.

The following in-plane deformations have been applied to the equilibrium honeycomb lattice under plane-strain border conditions:\cite{cadelano} (i) an uniaxial deformation $\zeta$ along the armchair direction, corresponding to a strain tensor $\epsilon_{ij}^{(ac)}=\zeta\delta_{ix}\delta_{jx}$; (ii) an uniaxial deformation $\zeta$ along the zig-zag direction, corresponding to a strain tensor $\epsilon_{ij}^{(zz)}=\zeta\delta_{iy}\delta_{jy}$; (iii) an hydrostatic planar deformation $\zeta$, corresponding to the strain tensor $\epsilon_{ij}^{(p)}=\zeta\delta_{ij}$; (iv)  a shear deformation $\zeta$, corresponding to an in-plain strain tensor $\epsilon_{ij}^{(s)}=~\zeta \left(\delta_{ix}\delta_{jy}+\delta_{iy}\delta_{jx} \right)$. 

In order to extend the reliability of the present model to electronic features under strain, our results about the effects of hydrostatic and uniaxial deformations on the band structure are at first compared with previous data  available  in literature. 
For graphene under in-plane hydrostatic deformation with $\zeta \leq 15\%$, both in compression and in traction, we have calculated the band electronic structure and the density of states. Since the hydrostatic strain does not change the $D_{6h}(6/mmm)$ symmetry of the hexagonal lattice (Fig.\ref{fig:structure+bz}a), we only observe the variation of the pseudogaps at $\Gamma$ and $M$ points, while the location of the Dirac points is clamped at the $K$ point.
In particular, the pseudogap at $M$ decreases almost linearly from 6 eV (for $\zeta= -15\%$) to 
1.8 eV (for $\zeta= +15\%$). We remind that its value for the unstrained configuration is $2.2$ eV.
These results are in quantitative good agreement with Ref.\onlinecite{gui}.
Any other non-hydrostatic deformation lowers the symmetry of the graphene lattice. When an uniaxial strain is applied, all the 6- and 3-fold rotational symmetries are lost: a transition from the hexagonal  $D_{6h}(6/mmm)$ to the rhombic $D_{2h}(mmm)$ symmetry is observed (Fig.\ref{fig:structure+bz}b).
The irreducible part of the first Brillouin zone (BZ) is also affected by such deformations, since its original triangular shape (Fig.\ref{fig:structure+bz}a) is varied to the polygonal form represented in Fig.\ref{fig:structure+bz}b.
The top of the valence band and the bottom of the conduction band are shown in Fig.\ref{fig:idro+uniax} for the undeformed configuration (panel a), as well as under uniaxial deformation (panels b and c, corresponding to a strain $\zeta=~15\%$ along the armchair direction and in the zig-zag direction respectively).
The main effect of strain is the opening of a pseudo-gap at K and K'. 
Accordingly, the Dirac points are no more located at such high-symmetry points; rather, they drift away within the BZ, either for deformations along armchair direction or along zig-zag one.
Once again this important qualitative feature is in good agreement with Ref.~\onlinecite{pereira:prb2009}.

\begin{figure}[t]
\centering%
{\includegraphics[width= 0.35\textwidth]{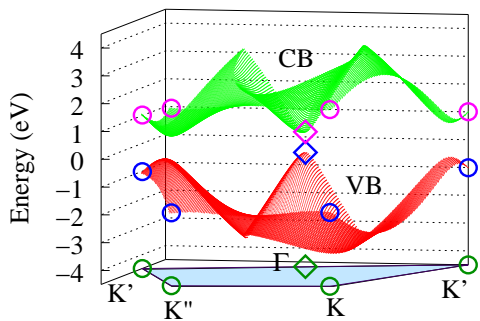}}\\
\caption{(Color online)
Top of the valence band (red, marked as VB) and bottom of the conduction band (green, marked as CB) of graphene under pure shear strain with
$\zeta=20\%$. Symbols connect the high-symmetry points of the BZ (bottom shaded area) to the energy of the corresponding electronic states.}
\label{fig:merge}
\end{figure}

Let us now consider the case of an in-plane shear deformation, described by the following shear strain
\begin{eqnarray}
\label{sheartensor}
  \hat{\epsilon}= \left( \begin{array}{cc}
0 & \zeta  \\
 \zeta & 0  \end{array} \right)
\end{eqnarray}
where $\zeta$ is the strain parameter. Such a deformation modifies the original reciprocal lattice vectors $\vec{b}_{1}$ and $\vec{b}_{2}$ into
\begin{eqnarray}
\nonumber
\vec{b}'_{1}&=& \frac{2\pi}{a_{\circ}}\left( 1+\zeta^{2} \right)^{-1}\left( \frac{1}{3}-\frac{\sqrt{3}}{3}\zeta, ~\frac{\sqrt{3}}{3}-\frac{1}{3}\zeta  \right)\\
\vec{b}'_{2}&=& \frac{2\pi}{a_{\circ}}\left( 1+\zeta^{2} \right)^{-1}\left( -\frac{1}{3}-\frac{\sqrt{3}}{3}\zeta, ~-\frac{\sqrt{3}}{3}-\frac{1}{3}\zeta  \right)
\end{eqnarray}

By applying the shear strain given in Eq.\ref{sheartensor} to the graphene lattice, its symmetry class is further lowered to monoclinic. The corresponding symmetry group is $2/m$.
Because of this change in symmetry, the irreducible part of the BZ is affected accordingly as shown 
in Fig.\ref{fig:structure+bz}c, which has K', K, K", K' as corners. In the undeformed lattice, a Dirac point is located at each of these corners. 
The scenario under shear strain is quite different from the case of uniaxial deformations: at the comparatively small strain $\zeta\simeq 16\%$, a gap is indeed opened.
The rise of a gap in the electronic band structure under shear is due to a peculiar process that involves the merging\cite{nota1} of two Dirac points, namely D' and D", which move away from the corners K' and $K"$ and approach each other inside the BZ. 

Is important to remark that the merging of the two inequivalent Dirac points and the opening of a gap, appear for a shear strain value  $\zeta\simeq 16\%$ which is lower than in the case of zig-zag uniaxial deformation.\cite{pereira:prb2009} The gap increases up to  a maximum value of $~0.72$ eV for shear strain parameter of $\zeta \simeq 20 \%$, as shown in Fig.\ref{fig:merge}.
We conclude that shear strain seems a likely candidate to achieve gap opening  in graphene for a deformation far enough from failure strain and, therefore, achievable with no danger for the overall mechanical stability of the two-dimensional sheet.

Gap opening is predicted by the present TB calculation to occur at an even smaller strain parameter $\zeta$, provided that a combination of shear and uniaxial strain is considered.
By adding an uniaxial component to shear we generate a strain tensor of the form
\begin{equation}
 \hat{\epsilon}= \left( \begin{array}{cc}
\zeta & \zeta  \\
 \zeta & 0  \end{array} \right)
~~~\text{or} ~~~\hat{\epsilon}= \left( \begin{array}{cc}
0 & \zeta  \\
\zeta & \zeta  \end{array} \right)
\end{equation}
for which the symmetry class of the lattice is not changed with respect to the pure shear case.

Nevertheless, uniaxial deformations along the armchair or zig-zag direction are found to dissimilarly affect the band structure of graphene. 
Only in the last case we have observed the merging of the Dirac points already at $\zeta\simeq 12\%$. The main features of the transition is the same as described before. The energy gap grows up to a maximum value of $~0.95$ eV (when the strain parameter achieves a value of $\zeta \simeq 17\%$), reducing again to zero at $\zeta \simeq 20\%$ due to the steady decrease of the direct gap at $\Gamma$. 

\begin{figure}[b]
\centering%
\subfigure[\label{dos shear}pure shear deformation]%
{\includegraphics[width= 0.35\textwidth, angle=0]{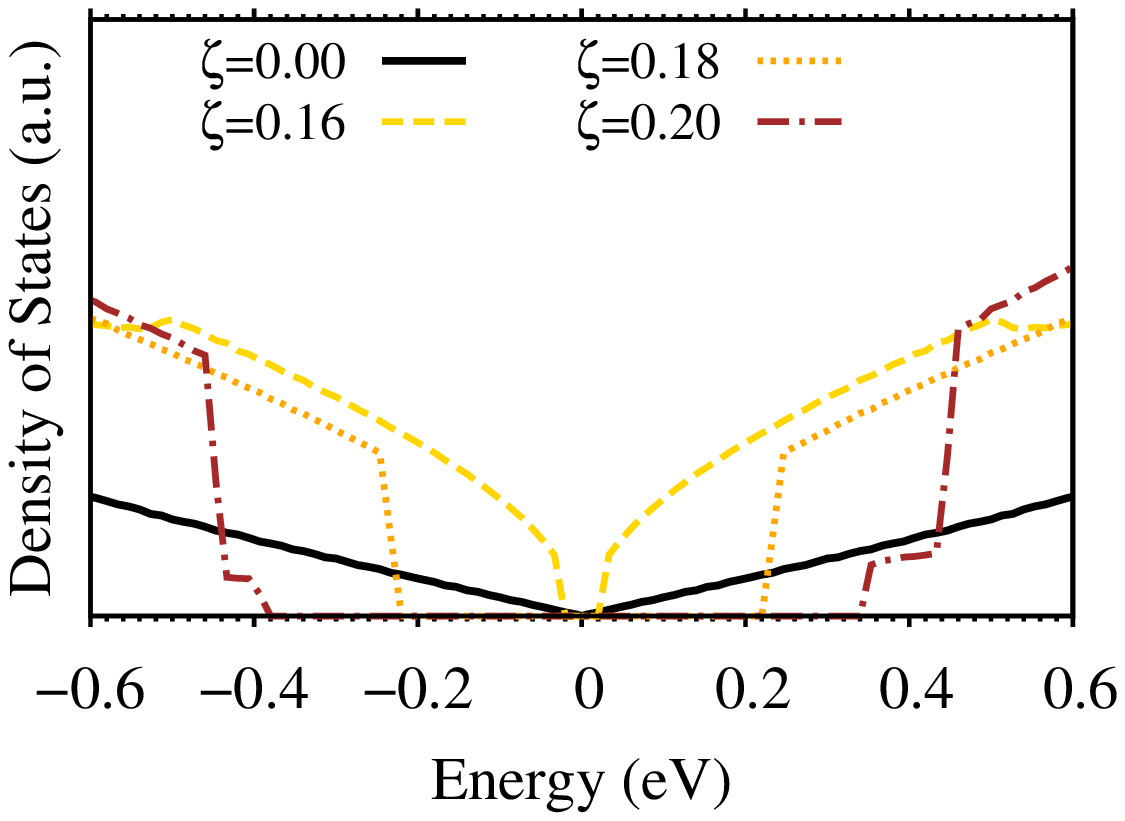}}\qquad
\subfigure[\label{dos shear plus ac}(shear $+$ armchair uniaxial) deformation]%
{\includegraphics[width= 0.35\textwidth, angle=0]{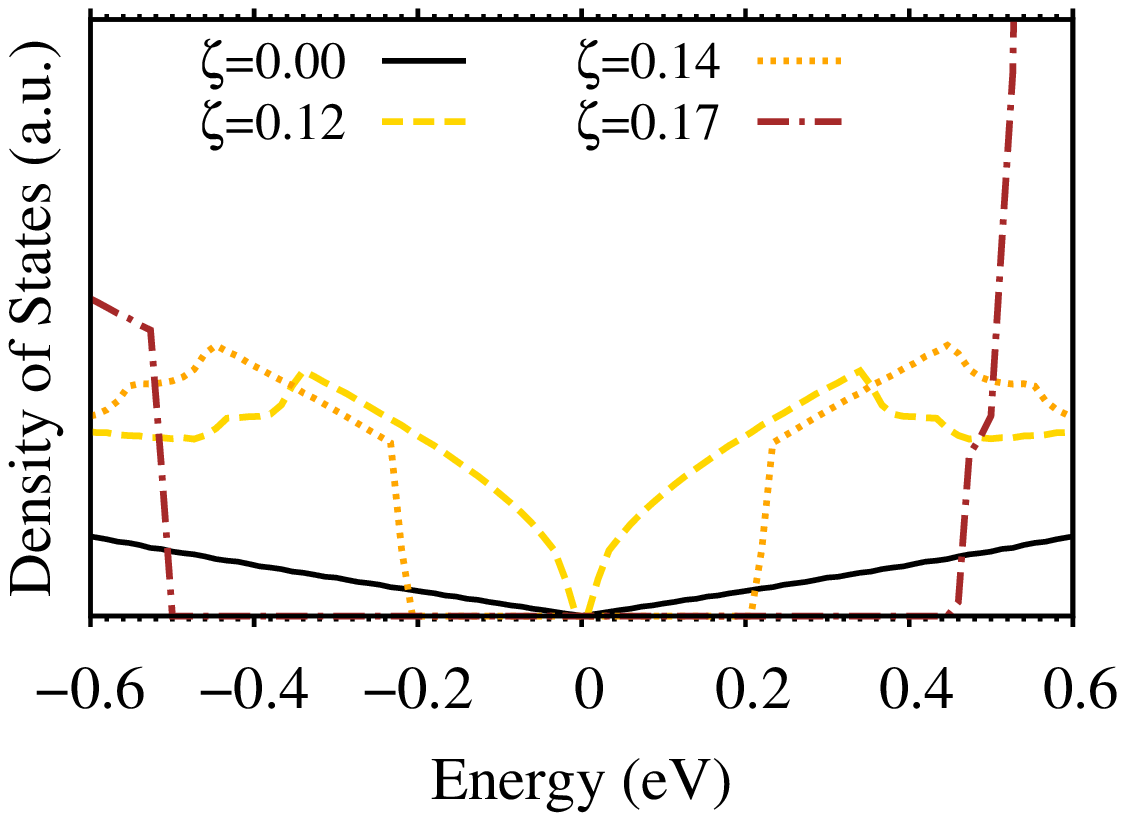}}
\caption{\label{capt:dos}(Color online) Density of states around the Fermi level (set conventionally at 0 eV) as function of the strain parameter   $\zeta$ . Panel (a): graphene under pure shear deformation. Panel (b): graphene under combined shear and uniaxial deformation (along the armchair direction). The maximum value of the energy gap is observed for a strain parameter as large as $\zeta \simeq 20\%$ and $\zeta \simeq 17\%$ respectively.}
\label{fig:dos}
\end{figure}

In order to quantitatively describe the evolution of the gap opening as function of the applied strain, the density of states (DOS) has been calculated by a two-dimentional $75\times 150\times 1$ regular $k$-point mesh of the (deformed) BZ.
As shown in Fig.\ref{fig:dos}, for a strain value less then $15\%$ (panel a) or $11\%$ (panel b), the DOS depends linearly on energy close to the Fermi level, showing a slope increasing with the strain. 
The two characteristic Van Hove singularities into the DOS move closer the Fermi energy and  disappear into abrupt gap-edges as soon as the gap is open. After the annihilation of the Dirac points, the DOS shows a  $\sim\sqrt{E}$ behavior.

\begin{figure*}[t]
\centering%
\subfigure[\label{capt2:HOMO LUMO shear +ZZ 15pc}(shear $+$ zig-zag uniaxial) deformation]%
{\includegraphics[width= 0.35\textwidth]{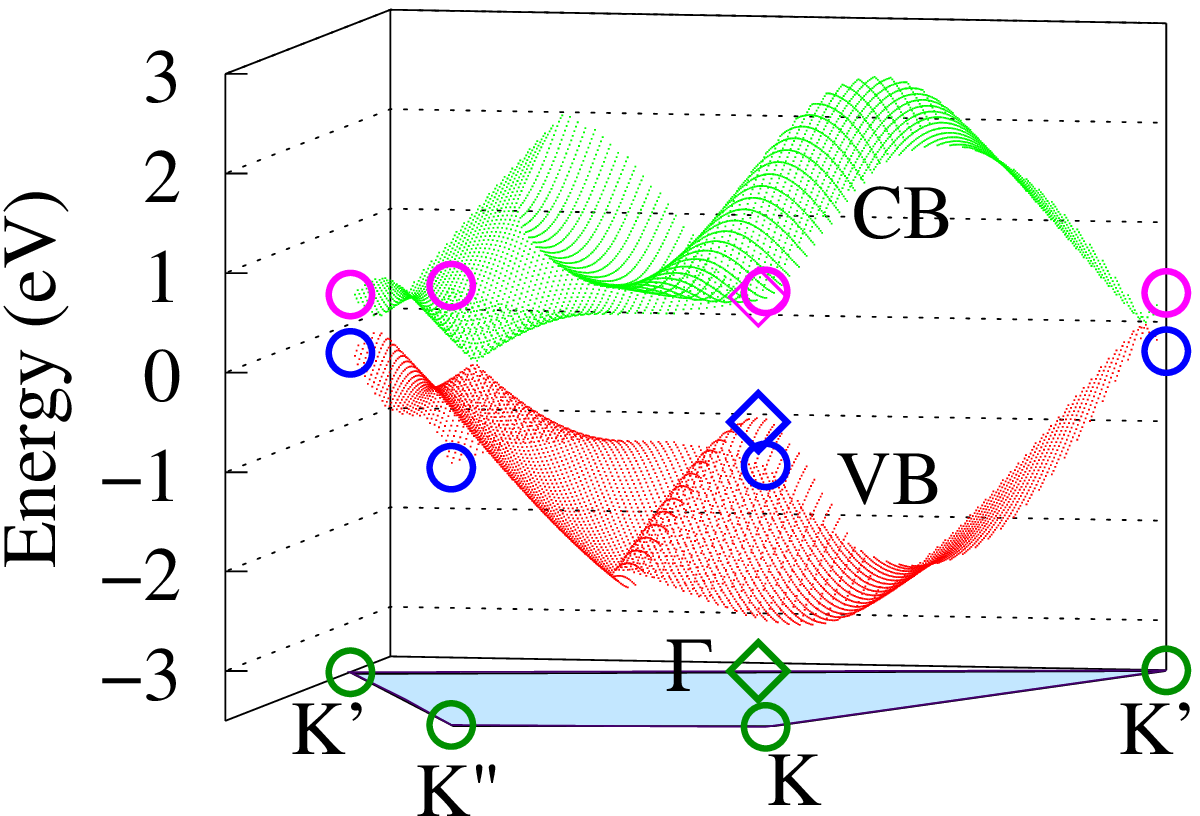}}\qquad
\subfigure[\label{capt2:HOMO LUMO shear +AC  15pc}(shear $+$ armchair uniaxial) deformation]%
{\includegraphics[width= 0.35\textwidth]{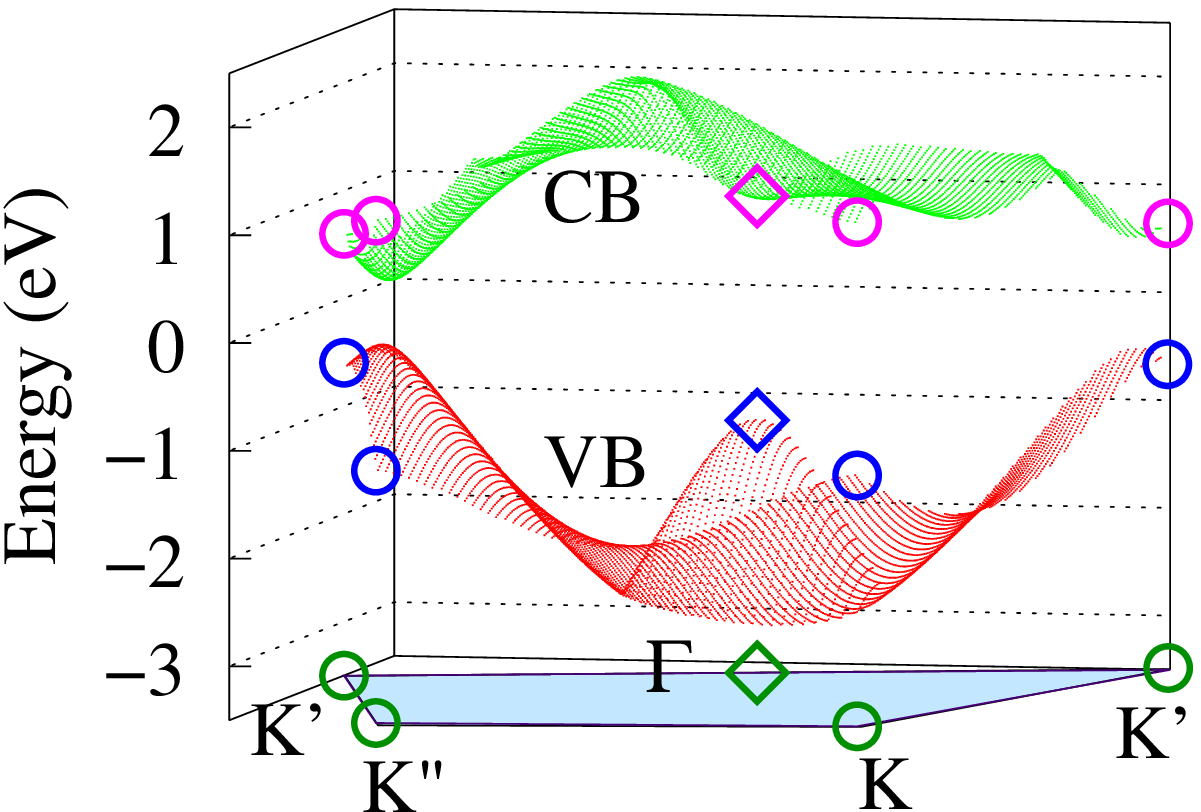}}%
\caption{(Color online)
Top of the valence band (red, marked as VB) and bottom of the conduction band (green, marked as CB) of graphene under combined shear and uniaxial strain with $\zeta=15\%$.  The uniaxial component of the strain is applied along the zig-zag (panel a) and armchair (panel b) direction. Symbols connect the high-symmetry points of the BZ (bottom shaded area) to the energy of the corresponding electronic states.}
\label{fig:shear}
\end{figure*}

We conclude by remarking that the two strain contributions (i.e. uniaxial and shear) could be combined in different ways so as to modulate the energy gap value.
In Fig.\ref{fig:shear}, the electronic band structures of graphene under different combinations of shear and uniaxial strain are compared, keeping the same value of the strain parameter $\zeta=~15$\%. While the combination of shear with uniaxial armchair shows a sizable energy gap of about $ 0.6$ eV, the combination of shear with uniaxial zig-zag is associated to a gapless band structure.

In conclusion, we have shown how the opening gap in the electronic spectrum of graphene could be achieved by applying deformations with a nonzero shear component, rather than a simple uniaxial deformation. Energy gaps have been found to vary in between $0.6$ eV and $0.9$ eV for a strain value far enough from  failure. In particular, we have shown that the most effective way to control the gap opening is to combine a shear and an armchair uniaxial deformation.

 We acknowledge computational support by COSMOLAB (Cagliari, Italy) and CASPUR (Rome, Italy).
Discussions with G. Fadda are gratefully acknowledged. One of us (L.C.) acknowledges partial financial support by the  
MATHMAT project (Universit\`a di Padova, Italy).

\bibliographystyle{apsrev}

\begin{thebibliography}{20}
\bibitem{nota1}
By merging we mean the following: at a critical  strain $\zeta\simeq 15.95\%$ the Dirac points are so close that they annihilate in a single hybrid Dirac cone, which shows a peculiar energy-momentum dispersion: it is linear (quadratic) along (perpendicular) to the direction joining the two Dirac points. At $\zeta=16.0$\%, a gap as small as 0.05 eV is eventually opened. More details about the motion of Dirac points in two-dimensional crystals under uniaxial stress are reported in Ref.~\onlinecite{montambaux, dietl,hasegawa}.
\end{thebibliography}

\end{document}